\documentclass[final,authoryear,5p,times,twocolumn]{elsarticle}
\usepackage{newtxtext,newtxmath}
\usepackage[T1]{fontenc}
\usepackage{ae,aecompl}

\usepackage{graphicx}	
\usepackage{amsmath}	
\usepackage{amssymb}	
\input psfig.tex
\usepackage{graphicx}
\usepackage{natbib}
\usepackage{array}
\usepackage{graphics}
\usepackage{latexsym}
\usepackage{amssymb}
\usepackage{amsmath}
\usepackage{fancyhdr}
\usepackage{morefloats}
\usepackage{bm}
\usepackage{color}
\usepackage{soul}

\newcommand\apjl{\@eapj@ApJLetters }

\title{Theory of multiple-stellar population synthesis in a non-Hamiltonian setting}

\author{S. Pasetto,$^{1}$
	D. Crnojevi\'c,$^{2}$
	G. Busso,$^{3}$
	C. Chiosi,$^{4}$
	L. P. Cassar\`{a},${^5}$
	\\
	$^{1}$The Observatories of the Carnegie Institution for Science, 813 Santa Barbara St., Pasadena, CA 91101, United States of America\\
	$^{2}$Department of Physics \& Astronomy, Texas Tech University, Box 41051, Lubbock, TX 79409-1051, United States of America\\
	$^{3}$Institute of Astronomy, University of Cambridge, Madingley Road, Cambridge CB3 0HA, United Kingdom\\
	$^{4}$Department of Physics \& Astronomy, "Galileo Galilei", University of Padua, Vicolo dell'Osservatorio 2, Padua, Italy\\
	$^{5}$ INAF-IASF,  Milano Via E. Bassini 15,  Milano, Italy
}

\date{Accepted XXX. Received YYY; }

\begin{document}
\begin{frontmatter}
\begin{abstract}
	We aim to investigate the connections existing between the density profiles of the stellar populations used to define a gravitationally bound stellar system and their star formation history: we do this by developing a general framework accounting for both classical stellar population theory and classical stellar dynamics.
 	We extend the work of Pasetto et al. (2012) on a single composite-stellar population (CSP) to multiple CSPs, including also a phase-space description of the CSP concept. In this framework, we use the concept of distribution function to define the CSP in terms of mass, metallicity, and phase-space in a suitable space of existence $\mathbb{E}$ of the CSP.
	
We introduce the concept of foliation of $\mathbb{E}$ to describe formally any CSP as sum of disjointed Simple Stellar Populations (SSP), with the aim to offer a more general formal setting to cast the equations of stellar populations theory and stellar dynamics theory. In doing so, we allow the CSP to be object of dissipation processes thus developing its dynamics in a general non-Hamiltonian framework.
	
Furthermore, we investigate the necessary and sufficient condition to realize a multiple CSP consistent with its mass-metallicity and phase-space distribution function over its temporal evolution, for a collisionless CSP. Finally, analytical and numerical examples show the potential of the result obtained. 	
\end{abstract}

\end{frontmatter}


\section{Introduction}\label{Intro}
Stars are the fundamental constituents of a galaxy. 
Our understanding of galactic structure and evolution depends very much on the processes governing their birth, and evolution. 
The evolutionary time scales of stars, their energy feedback, yields of chemically enriched material into the interstellar medium, end products of their evolutionary history, and distribution in space and time characterize the structure of the galaxies and govern their evolution. However, all these stellar phases and products are often subject to uncertainties of both theoretical and observational nature, generating a lacking comprehension of these important issues. The effort to address these difficulties must be carried on in a dual way: with the collection of new data and with the development of new theoretical frameworks to interpret these data. 
%

In the era of wide-field surveys, dealing with exponentially growing numbers of stars has become a challenge both for observational analyses and for their theoretical interpretation. In this contribution, we will address the latter.
The difficulties of dealing with a large number of stars have influenced historically both the classical stellar dynamics and the classic stellar population theories. In classical stellar dynamics, from the few-body problem the attention moved to the mathematical formulation of a many-body problem starting from the pioneering works of Eddington, Chandrasekhar, and others who applied the concepts of statistical mechanics (e.g., the Liouville and Boltzmann equation) and the theory of the potential to "groups of stars" subject to a shared gravitational potential and hence described by a distribution function \citep[e.g.,][]{2003gmbp.book.....H,1985gpsg.book.....S}. In the second half of the past century, a similar concept of "stellar populations" was used initially to address the fundamental equation of stellar statistics, the star-count equation \cite[e.g.,][]{1898PA......5..544S,1953stas.book.....T}. This concept reached the astronomy research field thanks to the observer W. Baade and finally proliferated in the Galaxy modeling field in the 80s \citep[see, e.g., ][]{1980ApJS...44...73B, 1984ApJS...55...67B, 1984ApJ...276..169B, 1985ApJS...59...63R}. In these works, the idea of stellar population involves the photometry alone without phase-space treatment \citep[e.g.,][]{1981ApJ...249...48G,1972A&A....20..383T,1973ApJ...186...35T}. The first works attempting a global model generalization can be dated back to \citet[][]{1987A&A...180...94B}, \citet[][]{1990ApJ...357..435C} and \citet[][]{1996AJ....112..655M}. 

We want to merge these two concepts of stellar populations coming from classical stellar dynamics theory and classical stellar population theory, with the goal to precisely define the minimum condition under which these theories give consistent results. On the one hand, classical stellar dynamics defines a composite stellar population by its density profiles: its natural environment is the phase-space where position and momentum determine the distribution of the stars in the phase-space. On the other hand, classical stellar population theory defines a composite stellar population through its star formation history and initial mass function: its natural environment is the mass and metallicity space within which the stars move according to the fuel consumption theorem.
To formulate a comprehensive framework  able to account for both the theories is a difficult mathematical task. Here we limit ourselves to the investigation of a simpler, but no less important, task that tightly connects to the star-count modeling techniques. We cast the problem in the following way: if both classical stellar dynamics and classical stellar population theories determine the total mass of a composite stellar population, which is the condition for these approaches to coincide? 
While for one composite stellar population the answer is known, this is not true for two or more stellar populations. In this work we will derive it for the first time \citep[see also][]{2018ApJ...860..120P}. 

In the literature, the concept of multiple stellar populations has long tradition and it is extensively used to study a large variety of topics \citep[e.g.,][]{1991AJ....102..951T,1995AJ....110.2105A,1996ApJ...469L..97A,1997AJ....114..680A,2001AJ....121.1013B,2002A&A...392...83B,2003AJ....125..770B,2005ARA&A..43..387G,2006A&A...451..125V,2008A&A...484..815B,2009ARA&A..47..371T,2010A&A...518A..43T,2012ApJ...744...58M,2014A&A...561A.141C,2014A&A...567A..46C} even if it still poorly defined or lacking mathematical formalism (see, e.g., \citealt[][]{2006essp.book.....S} or \citealt[][]{2011spug.book.....G} for a review on the subject). 

The most remarkable advancement in the mathematical treatment of groups of stars (i.e., populations) probably happened at the beginning of the past century with the introduction of continuous functions: although stars are discrete elements, large gravitationally bound groups of stars sharing common properties started to be studied using continuous distribution functions (DFs) and continuity relations, rather than set-theory (i.e., stars by stars summations). This represented a great advancement with respect to the Celestial mechanics punctual treatment based on the 3-body/few-body problem, etc.
In this work we introduce novel mathematical instruments, as the foliations, to address classical stellar population problems.

\citet[][]{2012A&A...545A..14P} introduced a new theoretical framework for the concept of stellar populations, and we here endorse and extend it to include multiple composite stellar populations (CSPs). This formalism has the advantage to include in the description of the classical dynamics of a CSP (based on the concept of distribution functions as well) the concepts that are natural to the theory of stellar populations (e.g., initial mass function, star formation rate, etc.). In the treatment that we are proposing, the star birth and death is formally included (hence changing the total number of stars) without any limitation on the nature of their dynamics. The formalism is correct both in the case of a collisional CSP of globular clusters, and a collisionless CSP of a galaxy. Furthermore, this formalism does not depend on the Hamiltonian nature of the dynamics  (see Sec. \ref{Sec4}).

The application of this general concept to the Milky Way (MW) has been presented in \citet[][]{2016MNRAS.461.2383P} and will be reviewed briefly in the next section. We start recalling some basic concepts and definitions from \citet[][]{2012A&A...545A..14P} and \citet[][]{2016MNRAS.461.2383P} in Sec. \ref{Sec21}. In Sec.\ref{Sec22} we set the basis for the idea of multiple stellar populations. In Sec.\ref{Sec23} we have a closer look at the necessary and sufficient condition for a system of the composite stellar population to be coherent in mass. In Sec.\ref{Sec3} we present two numerical examples which highlight the potential of the theory, in Sec.\ref{Sec4} we discuss our results and in Sec. \ref{Sec5} we draw our conclusions. The mathematical aspects of our work are detailed in Appendix A.

\section{Theory of multiple composite stellar populations}\label{Sec2}
\subsection{Basic concepts of a non-Hamiltonian statistical mechanics for CSPs}\label{Sec21}
A composite stellar population, or simply CSP, is a set of stars born at a different time $t$, positions $\bm{x}$, with different velocities $\bm{v}$, masses $M$, and chemical compositions $Z$. We assume that every star lives in the space $\mathbb{E}=M\times Z\times \bm{\varGamma }$ with $M\subset \mathbb{R}_{0}^{+}$ masses, $Z\subset \mathbb{R}_{0}^{+}$ metallicity, and ${\bm{\Gamma }} \equiv \left\{ {{{\bm{x}}^1},{{\bm{v}}^1},...,{{\bm{x}}^N},{{\bm{v}}^N}} \right\} \subset {\mathbb{R}^{6N}}$ phase-space ($N$ being the number of stars, and $\mathbb{R}_{0}^{+}$ the set of positive real numbers including zero)(\footnote{The choice of the domain of existence is arbitrary and made to exploit the following formalism. Other powerful solutions as $M\subset (\mathbb{R}_{0}^{+})^N$ for the space of masses, $[Fe/H]\subset \mathbb{R}^{N}$ for the space of metallicity, and $\bm{\varGamma }\subset {{\mathbb{R}}^{6N}}$ for the phase-space, can lead to a formalism in $\mathbb{E} \subseteq {(\mathbb{R}_{0}^{+})^N} \times {\mathbb{R}^N} \times {\mathbb{R}^{6N}}$ that is potentially interesting but more distant from classical stellar population theory.}). 
At each time $t$, a single realization of a CSP can be defined as a the $s^{th}$ set of points ${{E}_{s}} \in \mathbb{E}$ defined by some arbitrary properties (i.e., the variable of state of the CSP). Following classical statistical mechanics arguments, we consider not such a single realization of a CSP (microstate), but an infinite collection of the CSPs characterized by the same macroscopic state average (e.g., energy, density, velocity dispersion, metallicity, etc.) but different microscopic conditions, i.e., different microstates $s$. If a point ${{E}_{s}}$ is representative of the ${{s}^{\text{th}}}$-microstate we consider the set of all the $\left\{ s,q \right\}:{{E}_{s}}\ne {{E}_{q}}$ at any arbitrary $t$. Because the ensemble contains an infinite number of states, the change of the state variables of each CSP happens smoothly, i.e., continuously passing between neighboring states. This allows us to describe the CSPs by a distribution function ${{f}_{c}}:\mathbb{E} \to I\subset \mathbb{R}_{0}^{+}$ with $I$ finite interval of the real positive line including zero.  Under this hypothesis, the evolution of ${{f}_{c}}$ is given by the Liouville equation for non-Hamiltonian systems \citep[e.g.,][]{Thompson} that we write as:
\begin{equation}\label{(2.1)bis}
{\partial_t \left( {{g^{1/2}}{f}} \right)} + \left\langle {{\nabla _{\mathbf{x}}},{{{{\partial_t}} \mathbf x}}{g^{1/2}}{f}} \right\rangle  = 0,
\end{equation}
with $g(\mathbf x;t)$ being the metric tensor of $\mathbb{E}$ introduced above, which is the classic Liouville equation generalized to (non-Euclidean) dissipative spaces, as we assumed $\mathbb{E}$ to be. Hereafter $\nabla_\mathbf{x}$ refers to the gradient over a set of basis coordinates $\mathbf{x}$, $\left\langle { \bullet , \bullet } \right\rangle $ to the inner product, and $\partial_t$ to the partial derivative with respect to the time(\footnote{All these quantities exists because $\mathbb{E}$ is assumed to be a Riemannian manifold.}).

Every time a system presents irreversibility, e.g., the system presents dissipative processes, gas-processes, friction, interaction, merges, etc. it is non-Hamiltonian and non-Hamiltonian statistics has to be used to describe its irreversible dynamics. We can express the Eq.\eqref{(2.1)bis}  by introducing the evolution operator $\iota \mathcal{E}\left[ \mathbb{\bullet} \right]$:
\begin{equation}\label{(2.1)}
{{ \iota \partial_t {{f}_{c}} }}=\mathcal{E}\left[ {f_c} \right],
\end{equation}
with $\iota$ complex unit, $f_c \equiv \sqrt{g} f$ and $\frac{d}{dt}[{\bullet}]$ the total derivative operator(\footnote{The purpose of the multiplication by the imaginary constant is clearly to obtain an equation similar to the Schr\"{o}dinger equation, $\iota \hbar {\partial _t}\psi  = H\left[ \psi  \right]$, with $H[\bullet]$ the Hamiltonian operator, $2\pi \hbar $ Plank's constant, $\psi$ wave function, and to work with Hermitian operators (i.e. with real eigenvalues operators) even though we will not exploit here this features of $\mathcal{E}$.}). As mentioned above, in general the CSPs are non-Hamiltonian entities, and their total number of stars is not conserved. The only hypothesis that we require for Eq.\eqref{(2.1)} is that the DF is sufficiently smooth so that the necessary derivatives exist; we will assume for simplicity that ${{f}_{c}}\in {{C}^{\infty} (\mathbb{E})}$ (i.e., the set of continuous functions with infinitely continuous derivatives). The formal solution of Eq.\eqref{(2.1)} is then
\begin{equation}\label{(2.2)}
{{f}_{c}}\left( {{E} };t \right)={{e}^{-\iota \mathcal{E}t}}\left[ {{f}_{c}}\left( {{E} };0 \right) \right]={{e}^{-\iota \left( \mathcal{L}+\Lambda +\mathcal{F} \right)t}}\left[ {{f}_{c}}\left( {{E} };0 \right) \right],
\end{equation}
where, mutating the name from quantum mechanics, we call ${e}^{-\iota \mathcal{E}t}[{\bullet}]$ the evolution \textit{propagator}. Here we can decoupled the operator  $\mathcal{E}\left[ \bullet \right]$ linearly, in such a way that $\mathcal{E}\left[ \bullet \right]$ is split in a part granting the evolution and normalization of ${{f}_{c}}$ given by $\iota \mathcal{L}\left[ \bullet \right]$ (standard Liouville operator), and in a part accounting for the compressibility of $\mathbb{E}$ in the case of external fields, say $\iota \Lambda \left[ \bullet \right] $, whose function is to account for the compression of the phase-space without changing the number of stars. Finally, a third part, say $\iota \mathcal{F}\left[ \bullet \right]$, accounts for the rate of change of the number of stars in the stellar population, $\dot N = \dot N\left( t \right)$.

\subsection{Multiple stellar populations}\label{Sec22}
We will focus our attention on gravitationally bound systems of collisionless/collisional stellar populations, composed of ${{N}}={{N}}\left( t \right)<\infty $ stars as long as it is possible to identify unambiguously every star. We formally need only the enumerability of the stars, that for the purposes of the normalization of $f_c$ in $\mathbb{E}$ can be considered identical indistinguishable elements. (\footnote{The resulting distribution function is assumed to absorb the normalization factor accordingly (i.e., for the sake of simplicity, we omit cumbersome N! factors in the normalizations).}). We will ask also for a slightly more restrictive hypothesis of phase-space mixed CSPs in $\bm{\varGamma }$, with a detailed-balance in $Z$,  and non-interacting stars (e.g., we exclude interacting binaries). These hypotheses are in agreement with our request of continuity for the temporal evolution of $f_c$ and with the Liouville description introduced in Eq.\eqref{(2.1)} thus granting the possibility to be always able to disentangle two different CSPs in their evolution of time in the spirit of a non-Markovian evolution(\footnote{Violent relaxation, asymmetries and tidal forces can quench long range forces due to rapid changes in the gravitational potential. We will exclude from our consideration stochastic behaviors or a master-equation based approach.}).  In this way, we assume that it is always possible to know the position and velocity of each single star in $\mathbb{E}$, so that the concepts of distribution function in the phase-space and average metallicity of the stars are always well defined. This argument implies that a trajectory gives the evolution of a system in the extended ${\mathbb{E}}'\equiv \left( \mathbb{E};t \right)$ space, and two different initial conditions lead to distinct non-intersecting paths in ${\mathbb{E}}'$ called CSP \textit{orbits} in $\mathbb{E}$, $\mathbf{x}=\mathbf{x}(\mathbb{E};t)$ \footnote{Note how the notation $\mathbf{x}$ refers to the position of the state "s" in $\mathbb{E}$, while we save the notation $\bm{x}$ for the position of the $i^{th}$ star in the configuration space.}.

The equilibrium hypothesis for ${{f}_{c}}$ in the whole $\mathbb{E}$ does not hold strictly, i.e., there is not a globally defined ${{f}_{\infty}}$  that holds over the entire space $\mathbb{E}\forall t$ and to which the CSPs tend with increasing time. However, on limited-volume subsets of $\mathbb{E}$ and limited time intervals, we will be still able to define "stationary states" under a suitable hypothesis for the two-body relaxation time in $\bm{\varGamma }$ or time-independent main-sequence phases in $M\times Z$.  Hence, as expected by stellar dynamics and stellar structure standard theories, we will consider galaxies not as ergodic systems, but we will let isolating integrals to exist, and to foliate the phase-space $\bm{\varGamma }$ thus allowing us to speaking about, e.g., "families of orbits" in $\bm{\varGamma }$. In the same way, low-mass stars can live on main sequences with extremely long timescales where ${{f}_{\infty}}$  is virtually time-independent.

The theoretical framework developed in \citet[][]{2012A&A...545A..14P} for $\mathbb{E}$ found an application to the case of the Milky Way (MW) in \citet[][]{2016MNRAS.461.2383P}. We will not repeat it here, but we will focus on some aspects of the normalization with the intent of digging deeper into the constraints implied by such a formal approach. 

We need to state clearly two definitions(\footnote{We will leave the time dependence explicit in our equations as far as possible to develop our consideration in parallel with the original general formalism presented in \citet{2012A&A...545A..14P}. Moreover, although the Dirac notation is a winning one on the operators' algebra, we feel that the integral notation exploited in  \citet{2012A&A...545A..14P} is more common in this astrophysical context and we will keep using it here.}):

\textbf{Definition 1}: a simple stellar population (SSP) is a subset of $\mathbb{E}$ at constant $\bm{\varGamma }$ and $Z$.

\begin{figure}
	\includegraphics[width=\columnwidth]{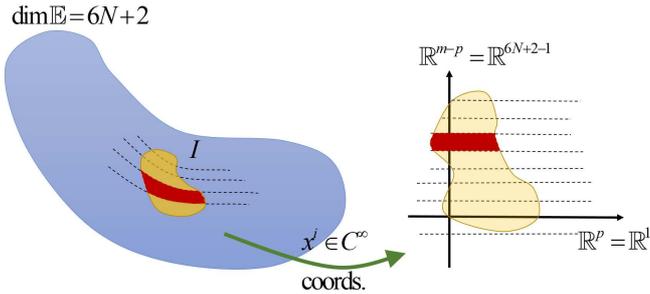}
	\caption{Foliation of the manifold $\mathbb{E}$ in leaves ${\mathfrak{F}}_{s}$. Each leaf is parallel and disjointed from any other leaf and covering the spectrum of masses, the co-dimension $p=1$ so that ${\mathbb{R}^p} = {\mathbb{R}^1} \equiv M$ while the left 6N-1 dimensions, i.e., the number of particles and the metallicity, are arbitrary but fixed, i.e., constant. Hence from the figure we can see that the section for $x^1 = M \ne \text{cnst.}$ is not constant, but $x^{1+1}=x^2 = Z = \text{cnst.}$ and so forth up to the section $x^{1+6N}$ are all constant by construction.} 
	\label{Folio}
\end{figure}

This represents the fundamental unit from which to construct the theory of CSPs. A collection of stars born at a given time $t$, with a single metallicity $Z$, and with a range of mass ${{\rm M}_{\text{SSP}}}\in \left[ {{\rm M}_{\min }},{{\rm M}_{\max }} \right]$ represents a line in $\mathbb{E}$ parallel to the $M$-axis. A set of these lines at the same $t$ for fixed $\bm{\varGamma }$, but spanning a range in metallicity $Z$, represents a CSP. Remembering that $\dim\bm{\varGamma }=6N$. it results natural to proceed with the following:

\textbf{Definition 2}: a 1-dimensional class ${{C}^{\infty }}$ foliation $\mathfrak{F}$ of the $6N+2$ dimensional differentiable manifold $\mathbb{E}$ (called space of existence) is a decomposition of the $\mathbb{E}$ into a union of disjointed connected SSPs (otherwise referred to as leaves ${{\mathfrak{F}}_{s}}$  of $\mathfrak{F}$ ), i.e., $\mathbb{E}=\coprod\limits_{{}}{{{\mathfrak{F}}_{s}}}$, with the following property: Every point $E\in \mathbb{E}$ has a neighborhood ${{I}}\subset \mathbb{E}$ and a system of local ${{C}^{\infty }}$ coordinates $\mathbf{x}=\left( M,Z,{{q}^{1}},...,{{q}^{3N}},{{v}^{1}},...,{{v}^{3N}} \right):I\to {{\mathbb{R}}^{6N+2}}$ such that for  $\forall {{\mathfrak{F}}_{s}}$ the components of $I\cap {{\mathfrak{F}}_{s}}$ are described by  the equations (see Fig.\ref{Folio}):
\begin{equation}\label{(2.3)}
\left\{ \begin{aligned}
& {\text{x}^{1+1}}=\text{cnst}\text{,} \\ 
& ... \\ 
& {\text{x}^{1+6N}}=\text{cnst.} \\ 
\end{aligned} \right.
\end{equation}

From the definition of a SSP, the mass function of a CSP at a given arbitrary  time $t$ reads \citep[see also Eq. (6) in ][]{2012A&A...545A..14P}:
\begin{equation}\label{(2.4)}
\begin{aligned}
M\left( t \right) &= \int_{}^{} {dMM} \int_{}^{} dZd\bm{\varGamma } {N{f_c}\left( {M,Z,{\bm{\varGamma }};t} \right) }  \hfill \\
&= \int_{}^{} {dMM} \sum\limits_{s \in \mathfrak{F}}^{} {\int_{}^{} {dZd{\bm{\varGamma }}N{f_c}\left( {M,Z,{\bm{\varGamma }};t} \right) \times } }  \hfill \\
&\times \delta \left( {{\bm{\varGamma }} - {{\bm{\Gamma }}_s},Z - {\text{Z}_s},M - {\text{M}_s}} \right) \hfill \\ 
\end{aligned} 	
\end{equation}
with $\delta \left( \bullet \right)$ being the multidimensional Dirac's delta and $\sum\limits_{s \in \mathfrak{F}}^{}\bullet$ the sum over all the SSPs. Integrals are supposed to extend over all the existence space unless stated otherwise. Eq.\eqref{(2.4)} is clearly equivalent:
\begin{equation}\label{(2.5)}
\begin{aligned}
M\left( t \right) = \int_{}^{} {dMM} \sum\limits_{s \in \mathfrak{F}}^{} {\int_{}^{} {dZd{\bm{\varGamma }}N{f_c}\left( {M,Z,{\bm{\varGamma }};0} \right) \times } }  \hfill \\
\times \delta \left( {{\bm{\varGamma }} - {{\bm{\varGamma }}_s}\left( t \right),Z - {Z_s}\left( t \right),M - {M_s}\left( t \right)} \right) \hfill \\ 
\end{aligned} 
\end{equation}
where $\bm{\varGamma }=\bm{\varGamma }\left( t \right)$ and $Z=Z\left( t \right)$ are the projections of the map $\mathbf{x}:{{\mathbb{R}}^{6N+1}}\to \mathbb{R}$, i.e., the orbit of $E$ in the section $\bm{\varGamma }$ and $Z$ of $\mathbb{E}$ with $\mathbf{x}=\left\{ \bm{\varGamma },Z \right\}$. In Eq.\eqref{(2.4)} the time dependence is due to the distribution function ${{f}_{c}}$, and the ${{\bullet}_{s}}$ quantities for the SSPs are integrated over $dZ d\bm{\varGamma }$. Eq.\eqref{(2.5)} transfers the time dependence to the phase variables being the two visions equivalents. The reader familiar with quantum mechanics can recognize in Eqs.\eqref{(2.4)} and \eqref{(2.5)} the equivalence between Schr\"{o}dinger representation and Heisenberg representation of the temporal evolution of the state-variables. In Eq.\eqref{(2.4)} the time dependence of the state variable (in our case the mass) is due to the distribution function ${{f}_{c}}$  (Schr\"{o}dinger representation), and the ${{\bullet}_{s}}$ quantities for the SSPs are integrated over $dZd\bm{\varGamma }$. Eq.\eqref{(2.5)} transfers the time dependence to the phase variables (Heisenberg representation). While foliating the space $\mathbb{E}$, the sum considered in Eq.\eqref{(2.4)} is weighted by $f_c$ that will add from $\mathfrak{F}$ only the non-null contributions. In the same way in Eq.\eqref{(2.5)}, and in the following Eq.\eqref{(2.7)}, the temporal evolution of each function accounts accordingly for the sum over the leaves of $\mathfrak{F}$.

We now can make use of the definition we made for the foliation of $\mathbb{E}$. We know that $\bm{\varGamma }=\bm{\varGamma }\left( t \right)$ is the solution of the Hamilton equations $\bm{\dot{\varGamma }}=\bm{\dot{\varGamma }}\left( t \right)$, and $\dot{Z}=\dot{Z}\left(t \right)$  is the chemical enrichment law of the population. Nevertheless, $\bm{\dot{\varGamma }}_s=0$  since $\bm{\varGamma }_s=\text{cnst}\text{.}$  as well as $\dot{Z_s}=0$  since $Z_s=\text{cnst}.$ $\forall {{\mathfrak{F}}_{s}}\in \mathfrak{F}$ in $\mathbb{E}$ (see Eq.\eqref{(2.3)}) because we foliated the space on SSPs, and by definition every SSP is a set of stars at a given position in the space $\bm{\varGamma }$ and at a  given metallicity in the space $Z$. Note that we need to have a formal expression for the equations of motion in $\bm{\varGamma }$ but we do not need them to come from an Hamiltonian vector field, e.g. by writing $\bm{\dot{\varGamma }} = J\frac{\partial H}{\partial \bm{\varGamma }}$ (with a $J$ standard symplectic matrix), nor we need any explicit form for the metallicity enrichment law $\dot{Z}=\dot{Z}\left( t \right)$. Furthermore, while for every leaf $\bm{\varGamma }=\text{cnst}\text{.}$ and $Z=\text{cnst}.$ (which is the way the ${\mathfrak{F}}_{s}$ are populated), in general it is $\dot{M}_s\left( t \right)\ne 0\forall s$ because in the same SSP the mass evolves with time following the fuel consumption theorem(\footnote{This is in general not true in the case of presence of close binaries (excluded for the purpose of this work) where the mass exchange through Roche lobes can play a major role in shaping the form of $M=M(t)$ and we loose the enumerability of the stars.}) \citep[][]{1981ApJ...249...48G}. Hence, Eq.\eqref{(2.5)} reads:
\begin{equation}\label{(2.6)}
\begin{aligned}
M\left( t \right) = \int_{}^{} {dMM} \sum\limits_{s \in \mathfrak{F}}^{} {N{f_s}\left( {M;0} \right)\delta \left( {M - {M_s}\left( t \right)} \right)}  \hfill \\
= \sum\limits_{s \in \mathfrak{F}}^{} {\int_{}^{} {dMM{{\hat \xi }_s}\left( {M;0} \right)} \delta \left( {M - {M_s}\left( t \right)} \right),}  \hfill \\ 
\end{aligned}	
\end{equation}
with $f_s$ referring to a DF for a SSP, while in the last line an initial mass function remains naturally defined as $\hat{\xi }_s\equiv Nf_s\left( M\left( t \right);0 \right)$ at each instant $t$. The shape of the mass function over each SSP ${{\mathfrak{F}}_{s}}$, is invariant, i.e., the mass function ${{\hat{\xi }}_{s}}$ for each SSP does not have an explicit temporal dependence, as expected from the concept of a SSP (Def. 1) which relies on a coeval set of stars. The rate of change of the mass with time $\forall s$ is specific to the SSP considered at the time envisaged for the SSP, and indeed it depends on the fuel consumption theorem \citep[][]{1981ApJ...249...48G} which allows us to write:  
\begin{equation}\label{(2.7)}
{{{\hat{\xi }}}_{c}}\left( M;t \right)=\sum\limits_{s\in \mathfrak{F}}^{{}}{{{{\hat{\xi }}}_{s}}\left( M;0 \right)\delta \left( M-{{\rm M}_{s}}\left( t \right) \right)}.
\end{equation}
For each SSP the contribution to the mass function is given by the product of a function that depends only on the mass, say an initial mass function (IMF) ${{{{\xi }}}_{c}}\left( M \right)$, times a function that for each mass depends only on the time, say the star formation rate (SFR) ${{\psi }_{c}}\left( t \right)$. We define them for the collective of the CSPs as:
\begin{equation}\label{(2.7bis)}
{{{\hat{\xi }}}_{c}}\left( M;t \right)\equiv {{{{\xi }}}_{c}}\left( M \right){{\psi }_{c}}\left( t \right).
\end{equation}
From the statistical point of view, Eq.\eqref{(2.7bis)} tells us that the probability to find a given star in the volume $dMdt$ and a second star in the volume $d{M}'d{t}'$ is just the product ${{\xi }_{c}}\left( M \right){{\psi }_{c}}\left( t \right)dMdt{{\xi }_{c}}\left( {{M}'} \right){{\psi }_{c}}\left( {{t}'} \right)d{M}'d{t}'$ without any correlation function.

This assumption is not strictly necessary to derive the total mass of a CSP, as we shall immediately see here below, but it is a commonly accepted assumption that follows from the supposed independence of the star formation processes over the time $t$(\footnote{Mathematically, the multiplicative separability is always possible if and only if (hereafter "iff") the functions are continuous and strictly positive (as in our case) because of the Kolmogorov Arnold "representation theorem".}). From Eq.\eqref{(2.6)} and Eq.\eqref{(2.7)} we obtain the total mass of a CSP (for a given interval of interest ${{\Delta t}}={{T}}-{{t}_{0}}>0$ referred to as the age of the CSP):
\begin{equation}\label{(2.8)}
\begin{aligned}
& {{\rm M}_{\text{tot}}}=\sum\limits_{c}^{{}}{\int_{{\rm M_{l}}}^{{\rm M_{u}}}{dM}M\int_{{{t}_{0}}}^{T}dt{{{{\hat{\xi }}}_{c}}\left( M;t \right)}} \\ 
& =\sum\limits_{c}^{{}}{\int_{{\rm M_{l}}}^{{\rm M_{u}}}{dM}M{{\xi }_{c}}\left( M \right)\int_{{{t}_{0}}}^{T}dt{{{\psi }_{c}}\left( t \right)}}.  
\end{aligned} 	
\end{equation}
with $\rm M_{l}$ and $\rm M_{u}$ being the lower and upper limits in the mass range considered in the CSP, and $\sum\limits_{c}^{{}} \bullet$ the sum over all the CSPs with at least a period of non-null SFR inside ${\Delta t}$.  

For example, let us suppose that we want to consider the different history of formation of the different CSPs in a completely isolated model of the MW (i.e., by considering CSPs of the galaxy as the Galactic bulge, disks, halo, spiral arms) in this unified formalism. While the total mass of the CSP ${{\rm M}_{\text{tot}}}$, is fixed because the system is assumed isolated, the contribution coming from the halo is built up much earlier in the history of the MW than the input to ${{\rm M}_{\text{tot}}}$ given by spiral arm populations, which contain predominantly young stars. Hence for ${{t}_{0}}=0$ and $T\cong 13.8\ \text{Gyr}$ (i.e., the age of the universe) the contribution in Eq.\eqref{(2.8)} from any CSP will be:
\begin{equation}\label{(2.9)}
{{\rm M}_{\text{tot}}}=\sum\limits_{c}^{{}}{\int_{{\rm M_{l}}}^{{\rm M_{u}}}{dMM{{\xi }_{0}}{{\Xi }_{c}}\left( M \right)\int_{{{t}_{0}}}^{T}{dt{{\psi }_{0,c}}{{\Psi }_{c}}\left( t \right)}}}, 	
\end{equation}
where, for reasons that will be clear soon, we wrote the IMF for a CSP as the product of a constant $\xi_0$ and a functional form $\Xi(M)$, i.e., $\xi(M) \equiv \xi_0 \Xi_c(M)$ (for some function $\Xi_c(M)$ here defined implicitly), and the star formation rate $\psi_c$ as the product of a constant $\psi_{0,c}$ times a functional form $\Psi_c(t)$ for every CSP, i.e., $\psi_c(t) \equiv \psi_{0,c} \Psi_c(t)$. The different CSPs that we need to consider have non-null SFR only within specific temporal intervals, e.g., the SFR of spiral arms in the MW can be chosen as a non-vanishing function only in the last 0.5 Gyr while the SFR of the MW halo is a non-null function only in the past 12-13 Gyr. For this reason it is convenient to write
\begin{equation}\label{(2.10)}
{{\Psi }_{c}}\equiv {{\tau }_{{{\left( {{t}_{\text{ini,c}}},{{t}_{\text{end,c}}} \right)}}}}{{\varphi }_{c}}\left( t \right),
\end{equation}
and ${{t}_{\text{ini,c}}}$ and ${{t}_{\text{end,c}}}\left( >{{t}_{\text{ini,c}}} \right)$ are the initial and final ages of star formation for the ${{c}^{\text{th}}}$ CSP considered. Here ${\varphi }_{c}$ is the functional form representing the SFR (e.g., an exponential, a linear profile, etc., see what follows) that we need to nullify outside a temporal interval of interest. We achieve this behavior with a "torii"-function (named after the traditional Japanese gates) i.e., a gate function that nullifies ${{\varphi }_{c}}\left( t \right)$ outside the limits ${{t}_{\text{ini}}}$ and ${{t}_{\text{end}}}$. One possible solution to achieve this functional feature is with a composition of Heaviside $\theta =\theta \left( z \right)$ functions as follows:
\begin{equation}\label{(2.11)}
\begin{aligned}
{\tau _{\left( {{\text{z}_1},{\text{z}_2}} \right)}}\left( \text{z} \right) &\equiv \theta \left( {\frac{1}{2}\frac{{{\text{z}_1} + {\text{z}_2} - 2\text{z}}}{{{\text{z}_1} - {\text{z}_2}}} + \frac{1}{2}} \right) -  \hfill \\
&- \theta \left( {\frac{1}{2}\frac{{{\text{z}_1} + {\text{z}_2} - 2\text{z}}}{{{\text{z}_1} - {\text{z}_2}}} - \frac{1}{2}} \right), \hfill \\ 
\end{aligned} 
\end{equation}
moreover, this torii function will be useful again later in a different context. For example, for the spiral arm CSPs, it can be assumed ${{t}_{\text{ini}}}=0.001\ \text{Gyr}$ and ${{t}_{\text{fin}}}=0.5\ \text{Gyr}$ since there is no evidence in the MW for the spiral arms to be as old as 13 Gyr. To obtain the total mass of the CSP, in Eq.\eqref{(2.9)} we assumed the normalization constant 
\begin{equation}
\begin{aligned}\label{ToryF}
&\int_{{\rm \rm M_l}}^{{\rm \rm M_u}} {\xi_c \left( M \right)MdM}  = \int_{{\rm \rm M_l}}^{{\rm \rm M_u}} {{\xi _0}{\Xi_c }\left( M \right)MdM}  \\ 
&= {\xi _0}\int_{{\rm \rm M_l}}^{{\rm \rm M_u}} {{\Xi_c }\left( M \right)MdM},  \\ 
\end{aligned}
\end{equation}
to be unique. Conversely, how the stars of a CSP evolve can differ from population to population and the normalization constants ${{\psi }_{0,c}}$ are left to vary from CSP to CSP, and we specify it with the individual index in Eq.\eqref{(2.10)}, $c=\left\{ \text{spr, hal, thn,}... \right\}$ for "spiral", "halo", "thin disks", respectively:
\begin{equation}
\int_{{t_1}}^{{t_2}} {\psi_c \left( t \right)dt}  = \int_{{t_1}}^{{t_2}} {{\psi _{0,c}}{\Psi_c }\left( t \right)dt}  = {\psi _{0,c}}\int_{{t_1}}^{{t_2}} {{\Psi_c }\left( t \right)dt},  
\end{equation}
where ${{\psi }_{0,c}}$ are constants to be determined as indicated in Sec.\ref{Sec23}.

We obtain now the same total mass of the CSPs by following a different path throughout the equations. We consider the propagator of $f_c$ as in Eq.\eqref{(2.2)}:
\begin{equation}\label{(2.12)}
\sum\limits_{c}^{{}}{\int_{{}}^{{}}{M{{f}_{c}}\left( M,Z,\bm{\varGamma };t \right)dMdZ}}=\sum\limits_{c}^{{}}{{{e}^{-\iota \left( \mathcal{L}+\mathcal{F}+\Lambda \right)t}}\left[ {{f}_{c}}\left( \bm{\varGamma };0 \right) \right]}.
\end{equation}
This equation can be simplified further under the assumption of the collisionless behavior of the stellar populations under examination. In the case we are interested to galaxies, the long two-body relaxation time hypothesis allows us to pass from a 6N degree of freedom description to a 6-dimensional phase-space with the volume elements $d\bm{\gamma }=\left( d\bm{x},d\bm{v} \right)$, by introducing the Boltzmann collisionless operator $\mathcal{B}$ to substitute the Liouville operator $\mathcal{L}$. In this simplified picture Eq.\eqref{(2.1)} becomes $\partial_t f_c =\left( \mathcal{B}+\mathcal{F} \right)\left[ f_c \right]$. For the Hamiltonian systems we then recover classical stellar dynamics results \citep[e.g.,][]{2014dyga.book.....B} for the density of the CSP as:
\begin{equation}\label{(2.13)}
\begin{aligned}
{{\rho }_{\text{tot}}}\left( \bm{x};t \right)&\equiv \int_{{{}}}^{{}}{d\bm{v}\sum\limits_{c}^{{}}{{{e}^{-\iota \left( \mathcal{B}+\mathcal{F} \right)t}}\left[ {{f}_{c}}\left( \bm{\gamma };0 \right) \right]}} \\ 
& =\sum\limits_{c}^{{}}{\int_{{}}^{{}}{d\bm{v}{{e}^{-\iota \left( \mathcal{B}+\mathcal{F} \right)t}}\left[ {{f}_{c}}\left( \bm{x},\bm{v};0 \right) \right]}},  
\end{aligned}	
\end{equation}
whose integral over the configuration space will yield the total mass at the instant considered. To account for the temporal evolution of the density profiles is not a trivial task, and as it often happens in stellar dynamics we are interested in a match with observations of ${{\rho }_{\text{tot}}}\left( \bm{x};t \right)$ at the present time, i.e., ${{\rho }_{\text{tot}}}\left( \bm{x};T \right)={{\rho }_{\text{tot}}}\left( \bm{x} \right)$. Hence, at the present time we can write:
\begin{equation}\label{(2.14)}
{{\rm M}_{\text{tot}}}=\sum\limits_{c}^{{}}{\int_{{}}^{{}}{d\bm{x}{{\rho }_{c}}\left( \bm{x} \right)}},
\end{equation} 
where ${{\rm M}_{c}} \equiv {\int_{{}}^{{}}{d\bm{x}{{\rho }_{c}}\left( \bm{x} \right)}}$ for every CSP, $c$.
This is the well-known relation for collisionless galaxy dynamics defined in $\bm{\varGamma }$, which is here obtained starting from the DF, ${{f}_{c}}$, defined in $\mathbb{E}$.  It represents the same quantity found in Eq.\eqref{(2.8)} starting from the same ${{f}_{c}}$ in $\mathbb{E}$ but obtained by following a different path through the equations. 

\textit{When are these mass determinations (from Eq.\eqref{(2.8)} and Eq.\eqref{(2.14)}) equivalent? What is the condition for the consistency of these two mass determinations? }In a mathematical formulation we can recast the question as follows: when does the relation 
$$\sum\limits_{c}^{{}}{\int_{{}}^{{}}{d\bm{x}{{\rho }_{c}}\left( \bm{x};t \right)}}=\sum\limits_{c}^{{}}{\int_{{}}^{{}}{dM{{\psi }_{c}}\left( t \right)M{{\xi }_{c}}\left( M \right)}}$$
hold at a given time $t$? When does this equation have at least a solution? Is it unique and how to determine it?

\textit{We move our new goal to the research of a consistent mass determination so that the total mass ${{\rm M}_{\text{tot}}}$ derived trough the standard stellar population theory, i.e. Eq.\eqref{(2.8)}, coincides with the total mass arising from the density profiles, Eq.\eqref{(2.14)}. }The only parameters left to be determined are the constants ${{\psi }_{0,c}}$ for multiple stellar populations. We show how to achieve this in the next section.

\subsection{A fundamental mass consistency condition for collisionless multi-stellar populations synthesis}\label{Sec23}
To answer the questions left in the previous section we proceed with the following definitions and by formulating the questions in the form of a theorem.
We define as \textit{consistent} a system of stars for which the following definition holds:

\textbf{Definition 3 [Consistent galaxy stellar population]}: Given a collisionless CSP identified by the distribution function ${{f}_{\text{c}}}\in I\subset \mathbb{R}_{0}^{+}$ in $\mathbb{E}\equiv M\times Z\times \bm{\gamma }$ ($I$ finite interval of the real positive line including the zero), for which the multiplicative separability of its mass function is given by ${{\hat{\xi }}_{c}}={{\xi }_{c}}\left( M \right){{\psi }_{c}}\left( t \right)$, it is said to be \textit{consistent} if it obeys to the fundamental relation:
\begin{equation}\label{(2.15)}
{\int_{{}}^{{}}{d\bm{x}{{\rho }_{c}}\left( \bm{x};t \right)}}={\int_{{}}^{{}}{dM{{\psi }_{c}}\left( t \right)M{{\xi }_{c}}\left( M \right)}}. 	
\end{equation}
Eq.\eqref{(2.15)} is a well-posed definition every time ${{f}_{c}}$ is non-negative (\footnote{This is a result sometime referred as Tonelli's theorem.}); a fact that always holds as a consequence of the definition of ${{f}_{c}}$ as a distribution function(\footnote{To make this definition explicit has also the aim to avoid nomenclature confusion with the concept of "dynamical consistency" used in stellar dynamical theory which means that given $ f_{\text{tot}} = \sum\limits_{c = 1}^{{N_p}} {{f_c}} $, Eq.\eqref{(2.1)bis} must hold with $f_c \geqslant 0 \forall c$ and ${\rho_{\text{tot}}} \equiv \sum\limits_{c = 1}^{N_p} {{\rho _c}}  = \sum\limits_{c = 1}^{N_p} {\int_{{}}^{} {{f_c}{d}{\mathbf{v}}} } $ and $\Delta {\Phi_{\text{tot}}} = - 4\pi G{\rho_{\text{tot}}}$.}). 

To understand when a set of CSPs can be said to be consistent for the case of a collisionless stellar system is the goal of the following theorem. 

\textbf{Theorem [Collisionless multiple stellar populations consistency theorem (MSP-CT)]}. Given a consistent composite stellar population (CSP) in the existence space $\mathbb{E}=M\times Z\times \bm{\gamma }$ defined by a DF ${{f}_{c}}\in I\subset \mathbb{R}_{0}^{+}$ ($I$ finite interval of the real positive line included the zero), we assume that multiplicative separability of CSP mass functions ${{\hat{\xi }}_{c}}$, i.e., ${{\hat{\xi }}_{c}}={{\xi }_{c}}\left( M \right){{\psi }_{c}}\left( t \right)$ holds,  where for every CSP ${{\xi }_{c}}\left( M \right)={{\xi }_{0}}\Xi_c \left( M \right)$ and ${{\psi }_{c}}={{\psi }_{0,c}}\Psi_c\left( t \right)$.
The CSP is consistent iff the system of equations:
\begin{equation}\label{(2.16)}
\begin{aligned}
&\sum\limits_c^{} {\int_{}^{} {d{\bm{x}}{\rho _c}\left( {{\bm{x}};T} \right)} }  =  \hfill \\
&= \sum\limits_c^{} {\int_{{\rm \rm M_l}}^{{\rm \rm M_u}} {dMM{\xi _0}{\Xi _c}\left( M \right)\int_{{t_0}}^T {dt{\psi _{0,c}}{\Psi _c}\left( t \right)} } }  \hfill \\ 
\end{aligned}
\end{equation}
\textit{has at least one solution}. In this case, the IMF normalization constant is
\begin{equation}\label{(2.17)}
{{\xi }_{0}}=\frac{{{\rm M}_{\text{tot}}}}{\sum\limits_{c}^{{}}{{{\psi }_{0,c}}{{I}_{\Psi ,c}}{{I}_{\Xi ,c}}}},
\end{equation}
with ${{I}_{\Xi ,c}}\equiv \int_{{\rm M_{l}}}^{{\rm M_{u}}}{dMM{{\Xi }_{c}}\left( M \right)}$ and 
${{I}_{\Psi ,c}}\equiv \int_{{{t}_{1}}}^{{{t}_{2}}}{dt{{\Psi }_{c}}\left( t \right)}$.  The SFR normalization constants are given by:
\begin{equation}\label{(2.18)}
{{\psi }_{0,c}}=\frac{{{\rm M}_{c}}\prod\limits_{j\ne c}^{{}}{{{I}_{\Xi ,j}}{{I}_{\Psi ,j}}}}{\sum\limits_{i}^{{}}{{{\rm M}_{i}}\prod\limits_{j\ne i}^{{}}{{{I}_{\Xi ,j}}{{I}_{\Psi ,j}}}}},
\end{equation}
where sums and products are assumed to run over all the CSPs.

\textbf{Proof}: The proof of Eq.\eqref{(2.16)} has been gradually achieved above with the passages from Eq.\eqref{(2.9)} through Eq.\eqref{(2.14)} once Defs. 1, 2, and 3 are considered. With the lemma in Appendix A, we conclude $\Box$.

Note that the collisionless nature of the CSP is an implicit hypothesis hidden in the definition of "consistency" of the CSP and it is necessary for the validity of the MSP-CT in the passage of Eq.\eqref{(2.13)}.

\subsection{Functional forms}
To fully exploit the theorem (a couple of examples will follow, and see also \citet{2018ApJ...860..120P}) we show a few self-standing results that are useful in handling the integrals in the previous theorem.  These results represent the "tools" to build up analytically a consistent set of CSPs once the MSP-CT is used. 

\subsubsection{Star-formation-rate profiles}
We will consider four star-formation-rate profiles:
\begin{enumerate}
	\item \textbf{Constant SFR}. We assume a constant star formation between two instants ${{t}_{2}}>{{t}_{1}}>0$: 
	\begin{equation}\label{(2.19)}
	\psi \left( t \right)={{\tau }_{\left( {{t}_{1}},{{t}_{2}} \right)}}{{\psi }_{0}}=\text{cnst.}
	\end{equation}
	identically. Considering ${{t}_{G}}>{{t}_{2}}>{{t}_{1}}>{{t}_{0}}>0$ and remembering that for the Heaviside theta function it holds the relation $\int_{{}}^{{}}{\theta \left( \text{z} \right)d\text{z}}=\text{z}\theta \left( \text{z} \right)+\text{cnst}\text{.}$, the integrals in Eq.\eqref{(2.18)} yield:
	\begin{equation}\label{(2.20)}
	{\psi _0}{I_\Psi } = {{\psi }_{0}}\int_{{{t}_{0}}}^{{{t}_{G}}}{dt\Psi \left( t \right)}={{\psi }_{0}}\left( {{t}_{2}}-{{t}_{1}} \right).
	\end{equation}
	
	\item \textbf{Exponential SFR}. We consider a profile 
	\begin{equation}\label{(2.21)}
	\psi \left( t \right)=\psi_0 \Psi(t)=\psi_0{{\tau }_{\left( {{t}_{1}},{{t}_{2}} \right)}}{{e}^{-\frac{t}{{{h}_{\tau }}}}},
	\end{equation}
	with ${{h}_{\tau }}\in \mathbb{R}\backslash \left\{ 0 \right\}$ non-null time scale length of an exponentially in/decreasing profile. We find for the integrals in Eq.\eqref{(2.18)} (with ${{t}_{G}}>{{t}_{2}}>{{t}_{1}}>{{t}_{0}}>0$):
	\begin{equation}\label{(2.22)}
	{\psi _0}{I_\Psi } ={\psi _0}\int_{{{t}_{0}}}^{{{t}_{G}}}dt{\Psi \left( t \right)}={{\psi }_{0}}{{h}_{\tau }}\left( {{e}^{-\frac{{{t}_{1}}}{{{h}_{\tau }}}}}-{{e}^{-\frac{{{t}_{2}}}{{{h}_{\tau }}}}} \right).
	\end{equation}
	
	\item \textbf{Linear SFR}. We investigate a linear pattern for the SFR between two assigned times, i.e., a shape
	\begin{equation}\label{(2.23)}
	\psi \left( t \right)={{\tau }_{\left( {{t}_{1}},{{t}_{2}} \right)}}{{\psi }_{0}}\left( \frac{{{\psi }_{{{t}_{2}}}}-{{\psi }_{{{t}_{1}}}}}{{{t}_{2}}-{{t}_{1}}} \right)\left( t-{{t}_{1}} \right)+{{\psi }_{{{t}_{1}}}}.
	\end{equation}
	Eq.\eqref{(2.18)} (with ${{t}_{2}}\ne {{t}_{1}}$, ${{\psi }_{{{t}_{2}}}},{{\psi }_{{{t}_{1}}}}$ all positive numbers) is then integrated entirely analytically (under the assumption of the previous case 1. and 2.) as:
	\begin{equation}\label{(2.24)}
	{\psi _0}{I_\Psi } =\frac{{{\psi }_{0}}}{2}\left( {{\psi }_{{{t}_{1}}}}+{{\psi }_{{{t}_{2}}}} \right)\left( {{t}_{2}}-{{t}_{1}} \right).
	\end{equation}
	It is clear that this kind of profiles once considered together with Eq.\eqref{(2.10)} can be combined to achieve any global SFR desired (see also Eq.\eqref{(2.10)}), where the age and metallicity relation are the result of a piecewise function.
	
	\item \textbf{Rosin-Rammler SFR}. Finally, it is of interest to present a SFR of the form (Rosin-Rammler, 1933):
	\begin{equation}\label{(2.25)}
	\psi \left( t \right)={{\tau }_{\left( {{t}_{1}},{{t}_{2}} \right)}}{{\psi }_{0}}{{t}^{\beta }}{{e}^{-\frac{t}{{{h}_{\tau }}}}},
	\end{equation}
	under the condition that ${{t}_{2}}>{{t}_{1}}>0$, $1\ne \beta >0$ is constant, and ${{h}_{\tau }}>1$ (see, e.g., \citealt[][]{1980A&A....83..206C}, \citealt[][]{2012A&A...548A..60G} for an extensive investigation of this family of profiles in relation to the MW chemical modeling or \citet{2012ceg..book.....M} for a review on the theory of chemical evolution of stellar populations). The integrals needed in Eq.\eqref{(2.18)} read:
	\begin{equation}\label{(2.26)}
	\begin{aligned}
	{\psi _0}{I_\Psi }  = {\psi _0}h_\tau ^{\beta  + 1}\left( {\gamma \left( {\beta  + 1,\frac{{{t_1}}}{{{h_\tau }}}} \right) - } \right. \hfill \\
	\left. { - \gamma \left( {\beta  + 1,\frac{{{t_2}}}{{{h_\tau }}}} \right)} \right), \hfill \\ 
	\end{aligned}
	\end{equation}
	where with $\gamma \left( a,z \right)=\int_{z}^{\infty }{dt{{e}^{t}}{{t}^{a-1}}}$ 
	we indicated the incomplete gamma function.
\end{enumerate}

\subsubsection{Mass function profiles}
We will consider three initial mass function profiles within preassigned mass limits $M\in 
\left[ {\rm M_{l}},{\rm M_{u}} \right]$:
\begin{enumerate}
	\item \textbf{Single power law IMF} \citep[e.g.,][]{1955ApJ...121..161S}. The integrals involved in Eq.\eqref{(2.16)}, with 
	\begin{equation}\label{(2.27)}
	\xi \left( M \right)={{\xi }_{0}}{{\Xi }}\left( M \right)={{\xi }_{0}}{{M}^{-\alpha }},
	\end{equation}
    and $\alpha =\text{cnst}\text{.}$ yield:
	\begin{equation}\label{(2.28)}
	{{\xi }_{0}}{I_{\Xi }}=\int_{{\rm M_{l}}}^{{\rm M_{u}}}{dMM{{\xi }_{0}}{{\Xi }}\left( M \right)}={{\xi }_{0}}\frac{\rm \rm M_{u}^{2-\alpha }-\rm M_{l}^{2-\alpha }}{\alpha -2}.
	\end{equation}
	
	\item\textbf{Piecewise linear functions }are very popular in the literature \citep[e.g.,][]{2001MNRAS.322..231K,1986FCPh...11....1S}. Hence it is worth to consider in detail what the normalization process required by the consistency theorems implies for these profiles. We require a single normalization factor for all the piecewise linear functions (${{\xi }_{0}}$ is unique as the total mass) so that continuity of the piecewise linear functions for different mass intervals requires a different scale function $\xi _{0}^{{{\alpha }_{i}}}$ for each mass interval, say $M\in \left[ {{\rm M}_{i}},{{\rm M}_{i+1}} \right[$: 
	\begin{equation}\label{(2.29)}
	\xi \left( M \right)=\sum\limits_{i=1}^{\rm N_{\text{sl}}}{{{\tau }_{\left( {{\rm M}_{i}},{{\rm M}_{i+1}} \right)}}{{\xi }_{0,{{\alpha }_{i}}}}{{M}^{-{{\alpha }_{i}}}}},
	\end{equation}
	where the function $\tau $ is the same previously introduced in Eq.\eqref{(2.11)} and $\rm N_{\text{sl}}$ is the number of slopes in the polygonal IMF considered. We need to determine the coefficients, ${{\xi }_{0,{{\alpha }_{i}}}}$, which grant continuity of the IMF in the points of connection of two consecutive slopes. Hence we solve the recurrence equation for the unknown generic coefficient ${{\xi }_{0,{{\alpha }_{i}}}}$ 
	\begin{equation}\label{(2.30)}
	{{\xi }_{0,{{\alpha }_{i}}}}=\rm \rm M_{i-1}^{{{\alpha }_{i}}-{{\alpha }_{i-1}}}{{\xi }_{0,{{\alpha }_{i-1}}}}\wedge {{\xi }_{0,{{\alpha }_{1}}}}={{\xi }_{0}},
	\end{equation}
	where we imposed an arbitrary condition for the global normalization in the first coefficient ${{\xi }_{0,{{\alpha }_{1}}}}={{\xi }_{0}}$ to the recurrence equation. The solution of the previous equation with this condition reads:
	\begin{equation}\label{(2.31)}
	{{\xi }_{0,{{\alpha }_{i}}}}={{\xi }_{0}}\prod\limits_{j=1}^{i-1}{\rm M_{j}^{{{\alpha }_{j+1}}-{{\alpha }_{j}}}}.
	\end{equation}
	Hence, in this way, we generalized the polygonal function with(\footnote{Note that fixed values of mass are referred to as $\rm M$ while the variable mass is referred to by the italic symbol $M$ throughout the paper.})
	\begin{equation}\label{(2.32)}
	\xi \left( M \right)={{\xi }_{0}}\sum\limits_{i=1}^{{{\rm N}_{\text{sl}}}}{\prod\limits_{j=1}^{i-1}{\rm M_{j}^{{{\alpha }_{j+1}}-{{\alpha }_{j}}}}{{\tau }_{\left( {{\rm M}_{i}},{{\rm M}_{i+1}} \right)}}{{M}^{-{{\alpha }_{i}}}}}.
	\end{equation}
	The most interesting case is for the number of slopes ${{\rm N}_{\text{sl}}}=3$, i.e., where $i=\left\{ {\rm M_{l}},{{\rm M}_{1}},{{\rm M}_{2}},{\rm M_{u}} \right\}$ are the lower mass, the first and second separation masses of the profile slopes, and the upper maximum mass considered respectively. In this case, for the integrals involved in Eq.\eqref{(2.16)} we get:
	\begin{equation}\label{(2.33)}
	\begin{aligned}
	&{\xi _0}{I_\Xi } = \int_{{\rm \rm M_l}}^{{\rm \rm M_u}} {dMM{\xi _0} \times }  \hfill \\
	&\times \sum\limits_{i = 1}^{3} {\prod\limits_{j = 1}^{i - 1} {\rm M_{j}^{{\alpha _{j + 1}} - {\alpha _j}}} {\tau _{\left( {{\rm M_i},{\rm M_{i + 1}}} \right)}}{M^{ - {\alpha _i}}}}  \hfill \\
	&= \sum\limits_{i = 1}^{3} {\prod\limits_{j = 1}^{i - 1} {\rm M_{j}^{{\alpha _{j + 1}} - {\alpha _j}}} {\tau _{\left( {{\rm M_i},{\rm M_{i + 1}}} \right)}}} \int_{{\rm \rm M_l}}^{{\rm \rm M_u}} {dMM{\xi _0}{M^{ - {\alpha _i}}}}  \hfill \\
	&= \sum\limits_{i = 1}^{3} {\prod\limits_{j = 1}^{i - 1} {\rm M_{j}^{{\alpha _{j + 1}} - {\alpha _j}}} {\tau _{\left( {{\rm M_i},{\rm M_{i + 1}}} \right)}}} \frac{{\rm M_u^{2 - {\alpha _i}} - \rm M_l^{2 - {\alpha _i}}}}{{\alpha  - 2}}, \hfill \\ 
	\end{aligned}
	\end{equation}
	where in the last line we made use of the Eq.\eqref{(2.28)}.
	\item \textbf{Lognormal IMFs} define a commonly used parametric family of profiles for stellar systems often used in combination with power-laws. We define them as:
	\begin{equation}\label{(2.34)}
	\xi \left( M \right)={{\tau }_{\left( {{\rm M}_{1}},{{\rm M}_{2}} \right)}}\frac{{{\xi }_{0}}{{C}_{a}}}{M}\exp {{\left( -\frac{1}{\sqrt{2}{{\sigma }_{\rm M}}}\log \frac{M}{{{\rm M}_{1}}} \right)}^{2}},
	\end{equation}
	in conjunction with power-laws as in \citet[][]{2003PASP..115..763C} or \citet[][]{1979ApJS...41..513M}. They can be equally fully integrated just noticing that the indefinite integrals hold for $M>{\rm M_{l}}$ 
	\begin{equation}\label{(2.35)}
	\begin{aligned}
	\int_{}^{} {dMM{\xi _0}{I_\Xi }\left( M \right)} & =  - {C_a}{{\text{e}}^{\frac{{\sigma _{\rm M}^2}}{2}}}{\rm \rm M_l}\sqrt {\frac{\pi }{2}} {\sigma _M} \times  \hfill        \\
	& \times {\text{erf}}\left(\frac{{\sigma _{\text{M}}^2 - \log M + \log {{\text{M}}_{\text{l}}}}}{{\sqrt 2 {\sigma _{\text{M}}}}} \right), \hfill
	\end{aligned}
	\end{equation}
	with $C_a$ and $\sigma_{M}$ as normalization constants, and $\rm erf(\bullet)$ is the Error function.
\end{enumerate}

These four profiles of the SFR and three of the IMF represent all the tools necessary to work with the previous theorem. With these fully analytical integrals at our hands, we can solve two numerical examples to show how the previous theorem acts. A sophisticated multi-stellar population model based on the MSP-CT is presented in \citet[][]{2016MNRAS.461.2383P} and \citet{2018arXiv180500486P}, and available on-line at www.galmod.org.

\section{Numerical tests}\label{Sec3}
\subsection{A simple model of the Milky Way potential}\label{Sec31}

\begin{table*}
    \centering
	\caption{Kinematic and dynamical properties of the MW components as derived after \citet{2016MNRAS.461.2383P}. Here we just mention that ${{M_B},{h_{r,B}}}$ are total bulge mass and radial scale length, $ {\rho _D,h_R,h_z,\Phi _0^a,h_{{\text{spr}}}^a,m,{\Omega _p},p,h_S}$ are central density, scale length, scale height, perturbation amplitude, spiral arm or bar scale length, total number of spiral arms, angular pattern speed, pitch angle, and shape function scale length, respectively, for all the disks exponential profiles and ISM. ${{\rho _{0,H*}},h_{r{H^*}},\alpha }$ are the stellar halo central density, scale length, and density slope, respectively, and $ {{v_0},h_{r,DM},q}$ are the scale velocity, scale length and flattening factor of the dark matter profile. Finally, ${{\bm{\sigma }}_{RR}}_ \odot $ is the only velocity dispersion tensor component necessary for the CSP considered along the principal axis of the system of reference of the population. }
	\label{Table1}
	\begin{tabular}{lllcc}
		\hline
		Components         & Scale parameters                                                                                                                                                                                & $\Delta t$                     & $\left[ {Fe/H} \right]$ & ${{\bm{\sigma }}_{RR}}_ \odot $  \\
		                   &                                                                                                                                                                                                 & $[\text{Gyr}]$                 &          [dex]          & [${\rm{km}}\;{{\rm{s}}^{ - 1}}$] \\ \hline
		                   & $\{{M_B},{h_{r,B}}\}$                                                                                                                                                                           &                                &                         &                                  \\
		                   & $\left[ {{{\rm{M}}_ \odot }\;{\rm{,kpc}}} \right]$                                                                                                                                              &                                &                         &                                  \\
		Bulge pop          & $9.3 \times {10^{9}},0.32$                                                                                                                                                                      & [6.0,12.0[                     &      [-0.40,+0.30[      &                                  \\
		                   &                                                                                                                                                                                                 &                                &                         &                                  \\
		                   & $  {\rho _D,h_R,h_z,\Phi _0^a,h_{{\text{sp}}}^a,m,{\Omega _p},t,p,h_S} $                                                                                                                        &                                &                         &                                  \\
		                   & $\left[ {{\rm{k}}{{\rm{m}}^{\rm{2}}}{{\rm{s}}^{{\rm{ - 2}}}}{\rm{kp}}{{\rm{c}}^{{\rm{ - 1}}}}{\rm{,kpc,km}}\;{{\rm{s}}^{{\rm{ - 1}}}}{\rm{kp}}{{\rm{c}}^{{\rm{ - 1}}}}{\rm{,deg,kpc}}} \right]$ &                                &                         &                                  \\
		Spr + Bar          & $9.47\times 10^6,2.00,0.17,887.82, 2.5, 2, 35.77, 0.13, 2.6$                                                                                                                                    & [0.1, 0.5[ $\bigcup$[5.0,12.0[ &      [-0.70, 0.05[      &               27.0               \\
		                   &                                                                                                                                                                                                 &                                &                         &                                  \\
		                   & $ \{{{\rho _D},{h_R},{h_z}}\}_\odot $                                                                                                                                                           &                                &                         &                                  \\
		                   & $ \left[ {{{\rm{M}}_ \odot }\;{\rm{kp}}{{\rm{c}}^{{\rm{ - 3}}}}{\rm{,kpc}}{\rm{,kpc}}} \right] $                                                                                                &                                &                         &                                  \\
		Thin disk          & $35.54\times 10^6,3.07,0.27$                                                                                                                                                                    & [0.5, 0.9[                     &      [-0.70, 0.05[      &               30.0               \\
		                   &                                                                                                                                                                                                 &                                &                         &                                  \\
		Thick disk         & $4.5\times 10^6,2.20,1.10$                                                                                                                                                                      & [10.0,12.0[                    &      [-1.90,-0.60[      &               51.0               \\
		                   &                                                                                                                                                                                                 &                                &                         &                                  \\
		ISM                & $22.63\times 10^6,4.51,0.20$                                                                                                                                                                    &                                &                         &                                  \\
		                   &                                                                                                                                                                                                 &                                &                         &                                  \\
		                   & $\{{{\rho _{0,H*}},h_{r{H^*}},\alpha }\} $                                                                                                                                                      &                                &                         &                                  \\
		                   & $ \left[ {{{\rm{M}}_ \odot }\;{\rm{kp}}{{\rm{c}}^{{\rm{ - 3}}}}{\rm{,kpc}}{\rm{,kpc}}} \right] $                                                                                                &                                &                         &                                  \\
		Stellar halo pop 1 & $4.9 \times {10^4},2.39,-2.44$                                                                                                                                                                  & [12.0,13.0[                    &        $<-1.90$         &              151.0               \\
		                   &                                                                                                                                                                                                 &                                &                         &                                  \\
		                   & $\{ {{v_0},h_{r,DM},q} \}$                                                                                                                                                                      &                                &                         &                                  \\
		                   & $\left[ {{\rm{km}}\;{{\rm{s}}^{-1}}{\rm{,kpc}}} \right]$                                                                                                                                        &                                &                         &                                  \\
		Dark matter        & $178.46,2.39,0.87 $                                                                                                                                                                             &                                &                         &                                  \\ \hline
	\end{tabular}
\end{table*}
We numerically test the validity of the Eqs.\eqref{(2.16)} involved in the MSP-CT. For this exercise, we choose to build up a simple MW potential (Table 1). This Table presents a perfectly functional MW model matching the major observational constraints on the MW potential. With  Table 1 and the density profiles in \citet[][]{2016MNRAS.461.2383P}, we obtain a total mass for the MW within 100 kpc of ${{\rm M}_{\text{100}}}\cong 0.8\times {{10}^{12}}\ {{\text{M}}_{\odot }}$,  the rotation curve at the solar location, ${{v}_{c}}\left( {{R}_{\odot }} \right)=228\ \text{km}\ {{\text{s}}^{-1}}$, the fraction of disk mass over the spiral component mass, $\frac{{{\rm M}_{\text{sp}}}}{{{\rm M}_{D}}}\cong 0.14$, the fraction of thick disk density over thin disk component, ${{\left. \frac{{{\rho }_{\text{thkD}}}}{{{\rho }_{\text{thnD}}}} \right|}_{\odot }}\cong 0.09$, the vertical force $\frac{{{F}_{z}}}{2\pi G}\left( {{R}_{\odot }},z=1.1\ \text{kpc} \right)\cong 69$ and $\frac{{{F}_{z}}}{2\pi G}\left( {{R}_{\odot }},z=2.0\ \text{kpc} \right)\cong 91$, and the Oort constants ${{O}^{+}}\left( {{R}_{\odot }} \right)=15\ \text{km}\ {{\text{s}}^{-1}}\ \text{kp}{{\text{c}}^{-1}}$ and ${{O}^{-}}\left( {{R}_{\odot }} \right)=-13\ \text{km}\ {{\text{s}}^{-1}}\ \text{kp}{{\text{c}}^{-1}}$(\footnote{It is beyond the goal of this paper to review the equations and the observational constraints considered in the MW potential. Nevertheless, in \citet{2016MNRAS.461.2383P} we presented the equations adopted to compute these values as well as a review of the most relevant observational constraints on these values.})

This benchmark model can be tested by setting these values in the on-line galaxy model web page (www.GalMod.org) of "GalMod" \citep[][]{2016MNRAS.461.2383P,2018arXiv180500486P} and setting to zero the density profiles of the thin disk population n$^o$ 3, 4 and 5. 

These parameters are obtained by minimizing a distance function in the parameter space from the best-values presented in \citet[][]{2016MNRAS.461.2383P}. The kinematic parameters presented in Table 1 and not involved in this exercise are left for completeness and were obtained by averaging the parameters in Table 2 in \citet[][]{2016MNRAS.461.2383P} \citep[i.e., they are not obtained from a direct fit of the data as for Table 2 in ][]{2016MNRAS.461.2383P}.

Taking the two CSP, e.g., spiral arm and thin disk, we can assume a constant star formation rate for the spiral arm over the first $t\in \left[ 0.1,0.9 \right]\ \text{Gyr}$. We exclude the first 100 Myr where we cannot properly speak of the ''stellar population'' because the stars are assumed to be still embedded in a collisional/dissipative environment, i.e., inside their parent molecular cloud or OB association locus. Note that the star formation rate is not inserted in $\left[ {{\text{M}}_{\odot }}\ \text{y}{{\text{r}}^{-1}} \right]$ because the role of the MSP-CT is to ensure the correct matching between the  amount of mass that results from the density profile parameters adopted (Col. 2 in  Table 1). For the star formation rate over the past 10 Gyr, $t\in \left[ 0.9,10.0 \right]\ \text{Gyr}$, we want to provide a more articulate profile for the SFR considering the significant temporal extension. We opt for Eq.\eqref{(2.25)} where we choose  $\left\{ \beta ,{{h}_{\tau }} \right\}=\left\{ 2.0,1.1\ \text{kpc} \right\}$ \citep[e.g.,][]{2010MNRAS.402..461J,2011MNRAS.411.2586J}. Because spiral arms came just from a perturbed distribution of a thin disk unperturbed mass distribution, both for the thin disk component and the spiral-arm component the total mass will be given by Eq.(45) in \citet[][]{2016MNRAS.461.2383P} for ${{R}_{\max }}\to +\infty $:
\begin{equation}\label{(3.1)}
{{\rm M}_{D}}=4\pi \sum\limits_{d=\text{spr,thn}}^{{}}{{{\rho }_{d}}h_{R,d}^{2}{{h}_{z,d}}{{e}^{\frac{{{R}_{\odot }}}{{{h}_{R,d}}}+\frac{{{z}_{\odot }}}{{{h}_{z,d}}}}}},
\end{equation}
which yields ${{\rm M}_{\text{spr}}}\cong 4.99\times {{10}^{9}}\ {{\text{M}}_{\odot }}$  and ${{\rm M}_{\text{thn}}}\cong 1.63\times {{10}^{10}}\ {{\text{M}}_{\odot }}$. From Eq.\eqref{(2.18)} we immediately get:
\begin{equation}\label{(3.2)}
\begin{aligned}
& {{\psi }_{0,\text{spr}}}=\frac{{{I}_{\Xi ,}}_{\text{thn}}{{I}_{\Psi }}_{\text{,thn}}{{\rm M}_{\text{spr}}}}{{{I}_{\Xi ,\text{thn}}}{{I}_{\Psi ,\text{thn}}}{{\rm M}_{\text{spr}}}+{{I}_{\Xi ,\text{spr}}}{{I}_{\Psi ,\text{spr}}}{{\rm M}_{\text{thn}}}}=0.354 \\ 
& {{\psi }_{0,\text{thn}}}=\frac{{{I}_{\Xi ,\text{spr}}}{{I}_{\Psi ,\text{spr}}}{{\rm M}_{thn}}}{{{I}_{\Xi ,\text{thn}}}{{I}_{\Psi ,\text{thn}}}{{\rm M}_{\text{spr}}}+{{I}_{\Xi ,\text{spr}}}{{I}_{\Psi ,\text{spr}}}{{\rm M}_{\text{thn}}}}=0.646, \\ 
\end{aligned}
\end{equation}
where just for the purpose of this example, ${{\Xi }_{\text{spr}}}$, ${{\Xi }_{\text{thn}}}$, ${{\Psi }_{\text{spr}}}$,  and ${{\Psi }_{\text{thn}}}$  are chosen to be  Eqs.\eqref{(2.28)}, \eqref{(2.33)}, \eqref{(2.20)} and \eqref{(2.26)} respectively, with IMF parameters from \citet[][]{1955ApJ...121..161S} and \citet[][]{2001MNRAS.322..231K}. Once we have obtained the normalization coefficient we can compute the number of stars that would fulfill the mass distribution $\rho $ for an IMF populated with masses in the interval $M\in \left[ {\rm M_{l}},{\rm M_{u}} \right]$. For the case considered above, we quickly obtain from Eq.\eqref{(2.17)} with Eq.\eqref{(3.2)} that
\begin{equation}\label{(3.2)bis} 
{{\xi }_{0}}=4.5828\times {{10}^{9}}\ {{\text{M}}_{\odot }},
\end{equation} 
so that the total number of stars for these two CSPs is
\begin{equation}\label{(3.3)}
\begin{aligned}
& {{N}_{\text{spr}}}=1.240\times {{10}^{10}} \\ 
& {{N}_{\text{thn}}}=3.618\times {{10}^{10}}. \\ 
\end{aligned}
\end{equation}
This result evidences the primary goal of the theorem: it adjusts the normalization functions so that the total amount of mass in stars realized by the star formation processes (with the assumed IMF) matches (at the instant considered) the total mass of the density profiles that generate the potential. Eq.\eqref{(2.18)} seems to be the only available option to compute algebraically the consistency condition for multiple stellar populations: even an algebraic software manipulator as Mathematica (Ver. 11.1.1) seems not to be able to produce algebraic solutions for N greater than two, giving output for ${{\psi }_{0}}$  that is at least a few pages long and virtually impossible to check and implement. Vice versa the explicit formulation presented in Eq.\eqref{(2.18)} allows us to easily handle many CSP's ${{\psi }_{0}}$  in a fully algebraic manner. Moreover, it allows the determination of the number of stars at the instant considered in concordance with the density profiles (and hence potential), IMF, and SFR. 

\subsection{Numerical solution of the star-count equation along any FoV}\label{Sec32}
\begin{figure}
\includegraphics[width=\columnwidth]{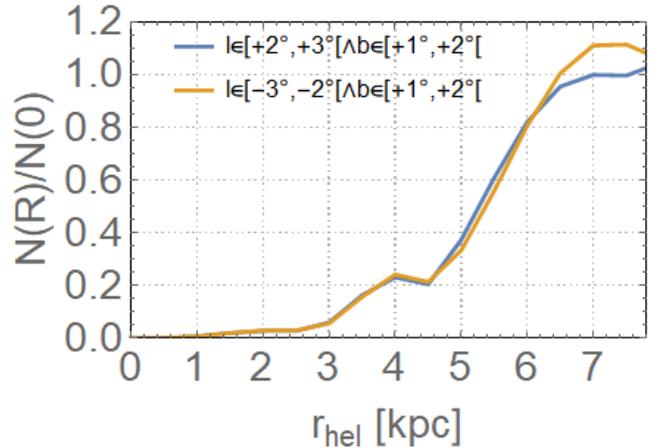}
	\caption{Relative number of stars distributed along the l.o.s in the selected direction. The solar location is at ${{r}_{hel}}=0$ corresponding to ${{\left\{ R,\phi ,z \right\}}_{\odot }}=\left\{ 8,0,0.02 \right\}\ \text{kpc}$. }
	\label{EvolTemp}
\end{figure}
To be able to determine the number of stars in a field of view $d\hat{\Omega }$ along any line of sight, it is of paramount importance to investigate the distribution of mass that generates it and, in turn, the underlying global gravitational potential. In this example, we show how to use the previous theorem to obtain this significant quantity. For the sake of this exercise, we will omit the observational color-magnitude diagrams investigation which assume it possible to observe all the stars within a mass range in the $\hat{\Omega }$ of interest. Differently from the previous exercise, we exploit here the configuration space dependence of the MSP-CT i.e.,  the left-hand side of Eq.\eqref{(2.16)}. We base this example again on the potential of  Table 1, and we solve Eq.\eqref{(2.16)} along an arbitrary but fixed direction. 

To test the central bulge/bar model decomposition \citep[whose details are introduced in ][]{2016MNRAS.461.2383P,2018arXiv180500486P}, we choose two small fields of view (FoV) in the direction $l\times b\in \left[ 2{}^\circ ,3{}^\circ  \right[\times \left[ 1{}^\circ ,2{}^\circ  \right[$ and $l\times b\in \left[ -3{}^\circ ,-2{}^\circ  \right[\times \left[ 1{}^\circ ,2{}^\circ  \right[$. We compute the same equations of the previous exercise. For the six stellar populations we evaluate six integrals ${I_{\psi ,i}}$ for $i=1,..,6$ from Eq.\eqref{(2.25)}, \eqref{(2.25)}, \eqref{(2.19)}, \eqref{(2.25)}, \eqref{(2.19)}, \eqref{(2.18)} for the bulge, bar, thin disk 1, thin disk 2, thick disk, and halo, respectively (with parameters as in Table 1). IMF profiles are taken from Eq.\eqref{(2.31)} and the integrals  ${I_{\Xi ,i}}$ for $i=1,..,6$ computed accordingly. The integrals of Eq.\eqref{(2.14)} were computed in cones of increasing size throughout the FoV directions.
In Figure 1 we plot the relative number of stars $N\left( R \right)$ normalized to the central galaxy value $N\left( 0 \right)$. The two FoVs start with the same number of stars in each FoV and the $N\left( R \right)$  trend is dominated by the "cone effect" of the opening angle. Nonetheless, already at about an heliocentric distance ${{r}_{hel}}\cong 4\ \text{kpc}$ the difference in the stellar density profiles due to the different direction starts to be visible. The non-axisymmetric effects are mostly visible around ${{r}_{hel}}=7\ \text{kpc}$ where the l.o.s. along the positive longitude meets the bar overdensity and grows while the negative longitude does not show the bar effect (further details in a dedicated paper, \citet{2018arXiv180500486P}). 

\section{Discussion}\label{Sec4}
Every time a system presents irreversibility (i.e., dissipative processes, gas driven processes, friction, mergers, etc.) then non-Hamiltonian statistics has to be used to describe its dynamics. The galaxies do not represent an exception. Their stellar component origins from gravitationally bound clouds of gas that evolve converting (in an irreversible way) the hot gas to molecular gas,  then to stars and again to chemically processed gas ejected into the ISM in an irreversible cycle of gradually increasing entropy (in an isolated system).
We described this continuous dynamic in a space $\mathbb{E}$ from the point of view of stellar populations, i.e. a set of discrete elements (stars) that are allowed to be created, to evolve and to die while moving in the space.
To realize a sounder mathematical setting for the concept of the stellar population, we made use of foliations in the existence space of the stellar populations in SSPs. 

This framework has the advantage of dealing with integral quantities instead of the discrete set theory (i.e., with distribution functions on well-behaved manifolds).  In \citet[][]{2016MNRAS.461.2383P} it has proven its enormous advantage by solving the classical star-count equation outside the "small FoV" framework. This resulted in a star count solution for a large FoV that is a particularly promising result when considering the ever increasing datasets stemming from current and upcoming partial or whole-sky surveys.

The second and more significant advantage inherited from the concept of the distribution function is the full incorporation of the dynamics in the stellar population treatment. In the existence space $\mathbb{E}$, the distribution function is treated with a mathematical formalism borrowed from quantum mechanics and the fundamental units are represented by a time-invariant element, with which the existence space  $\mathbb{E}$ can be foliated (i.e., SSPs). This offers an elegant theoretical formalism and a rich mathematical background from quantum mechanics to exploit. Finally, the extension of the concept of stellar populations to combine the classical stellar population theory and the stellar dynamics theory as presented here  aims to give a solid mathematical basis for the classical research on stellar systems as described in $\mathbb{E}$ \citep[e.g., pioneered by ][]{1987A&A...180...94B,1996AJ....112..655M,1986A&A...157...71R}. 

We note how our concept of foliation introduced in Sec.2.2 can be naturally pushed further to describe the phase-space of collisionless systems, say ${{\bm{\gamma }}}$, as a special case. If we consider a collisionless stellar system such as a galaxy, e.g., our MW, and we assume it to evolve in complete isolation (i.e., we exclude tidal interactions with the dwarf companions, complete phase-mixing in ${{\bm{\gamma }}}$, streams, star cluster inside the galaxy, binary interactions, etc.) we can try to exploit the Jeans theorem to foliate  ${{\bm{\gamma }}}$ by (isolating) integral of motions \citep[e.g.,][]{1962MNRAS.124....1L}.
In this case, we are able to write the DF for each SSP as ${f_{{\text{SSP}}}} = {f_{{\text{SSP}}}}\left( {M,Z,{I_{1,..,n}}} \right)$ for $I_i$ integrals of motions in ${{\bm{\gamma }}}$. 
Unfortunately the limitations in the applicability of this approach can be severe. The observations of the MW in particular show the existence of a bar in the MW center and spiral arms (i.e., non-inertial CSPs that slow down kinematic heating), accretion events (e.g. from dwarf galaxies) that apply torque to the MW angular momentum, etc. All these events induce a violation of the conservative (i.e. Hamiltonian) nature of the systems. Probably the most prominent example of a non-Hamiltonian, time-irreversible system is our own Galaxy. Modern research to overpass the limitations of the Hamiltonian (or action-based) formalism is lead by N-body numerical simulations \citep[e.g.,][]{2014MNRAS.445..175G,2003MNRAS.340..908K} or analytical studies \citep[e.g.,][]{2010A&A...522A..30C}.

Finally, we stress how MSP-CT not only offers \textit{consistency} in the existence space of the CSP, but it also provides the number of stars in an entirely general setting (without symmetry conditions on the underlying stellar populations). It sets the basis for the star count technique \citep[][]{2016MNRAS.461.2383P,2018arXiv180500486P}.
The theorem does not claim the uniqueness of the solution. The theorem is indeed the outproduct of an average procedure on all the possible microstates of the CSP, i.e., on an \textit{ensamble} as introduced in Sec.\ref{Sec2}. To unequivocally specify the microstate is beyond the framework of the theory and it would require the specification of the evolution operator $ \mathcal{E}$ in detail. In Eq.\eqref{(2.1)} we should give explicit formulation to the Liouville operator $\iota \mathcal{L}\left[ \bullet \right]$ thus relating $f_c$ with the total gravitational potential through the Poisson equation(\footnote{See footnote 10}), to the compression operator $\iota \Lambda \left[ \bullet \right] $ (to account for the presence of gas influencing the dynamics of the system or the presence of external systems) and finally to the $\iota \mathcal{F}\left[ \bullet \right]$ thus accounting for the change in the number of stars in agreement with the equation of stellar structure.
In particular, in the extended space $\mathbb{E}'$  for a given input physics (equation of state, nuclear reactions, opacity,  etc.), the mass and chemical composition of a star, the structure and hence the position on the Hertzsprung-Russell diagram (HRD) should be uniquely determined \citep[][]{2012sse..book.....K}. Unfortunately, the Vogt-Russell Theorem has never been proven on strict mathematical basis, and sometimes the presence of loops in intermediate mass stars at given input physics seems to have an erratic behavior: two stellar models with the same internal structure seem to correspond to two different locations on the HRD (one red and the other blue) thus resulting in a violation of the Vogt-Russel Theorem \citep[][]{1973A&A....27..323L,1972A&A....19..473L}. In relation to this, it is worth recalling that all stellar models are calculated with numerical methods so that the claim that two stellar models are identical is always hampered by this inherent drawback.
In any case,  several  decades of systematic applications of stellar evolution theory and their results (isochrones, synthetic HRDs,  etc.) to study stellar populations in clusters and fields, have always provided a consistent interpretation of the observational data. To conclude, we are inclined to consider the Vogt-Russell Theorem always verified and the stellar models in use "unique" even though further investigation is required.

This theorem is neither present, nor are there any similar consistency conditions implemented in any of the available star count models proposed in the literature, such as the Besan\c{c}on model \citep{2003A&A...409..523R}, Trilegal \citep{2005A&A...436..895G}.  Every time the gravitational potential is involved in the generation of the stellar kinematics, all the stellar populations must be considered simultaneously (because their combined gravitational potential enters in the collisionless Boltzmann equation). In this way, the fundamental theorem of mass consistency must be applied to generate the number of stars even when the model works in the small field of view approximation (i.e., whenever gradients of the density distribution are not relevant, see \citealt[][]{2016MNRAS.461.2383P}).

\section{Conclusions}\label{Sec5}
In this paper, we investigated the mass-consistency relation between the concept of stellar populations as intended by the classical stellar dynamical theory and the classical stellar population theory. This is done in the framework of a generalized concept of the stellar population developed by \citet{2012A&A...545A..14P}, where phase-space, mass, and metallicity of the stars are treated using distribution functions defined in a suitable existence space. We obtain a condition (expressed in the form of a theorem) that has immediate applications to a star-count model technique.

Traditionally, a population (here a stellar population) is considered as a set of elements (stars) sharing common properties(\footnote{In a more formal way, we could say that a relation of equivalence is partitioning the set through equivalence classes, with the stellar properties of interest defining the quotient set.}). Because of numerous stars typically involved, it is often more convenient to speak about distribution functions in a suitably defined manifold of existence for the stellar population. This framework is commonly adopted in stellar dynamics and was exploited for the first time in stellar populations by \citet[][]{2012A&A...545A..14P}.

When does the total stellar mass of the galaxy obtained by stellar dynamics theory equal the total stellar mass obtained by stellar population theory? In this work, we found a new answer in the form of the MSP-CT to this old question(\footnote{More correctly, MSP-CT not only aims to answer this question but also to generalize the answer to an arbitrary number $N<\infty$ of CSPs.}).  

We started exploring what we can learn by endorsing the \citet[][]{2012A&A...545A..14P} formalism. In particular, for the case of a galaxy (i.e., an approximatively collisionless-dynamics stellar system), we obtained a theorem (MSP-CT) that proves how the solution of the classical condition of equivalence between stellar dynamical mass $\text{M}^\text{dyn}$, and stellar population mass $\text{M}^\text{str}$, say $\text{M}^\text{dyn}=\text{M}^\text{str}$, always exists (it is not unique, as seen in the Lemma appendix A) and it is given by the set of Eqs.\eqref{(2.16)}, \eqref{(2.17)}, and \eqref{(2.18)}, obtained for the first time here.

Aside from the mathematical formalism, the physical interpretation of the MSP-CT can be understood as follows. The relative contribution to the total mass of two or more stellar populations depends at every instant on their relative density distribution according to their star formation history. The way in which the total mass is distributed among the stars (or between the different CSPs) depends on the SFH and the IMF of these populations. As time elapses, the stellar population ages and the stars leave the main sequence, die, or enter a quiescent stage (white dwarfs, neutron stars). During their life, they recycle material and enrich the interstellar gas violently (e.g., as supernovae) or quietly (as stellar winds). The metal abundances increase due to both self-enrichment by the parent SSP and the contribution from all other stellar SSPs. The way in which these stars are distributed at every time $t$ is determining their number and their overall mass in any arbitrary volume of the galaxy, a result that is given (in an analytical way) by the mass-consistency theorem for composite multiple-stellar populations (MSP-CT).

The applications of the MSP-CT are left to a dedicated paper that introduces GalMod and its features \citep[][]{2018arXiv180500486P}.

Furthermore, our formalism is not limited to galaxies. We worked out a framework that extends the applicability of the concept of multiple composite stellar populations to a large variety of gravitationally bound stellar systems.
The concepts we have developed in this paper can be applied straightforwardly to studies of globular clusters with single or multiple stellar populations \citep[e.g.,][]{2012ApJ...744...58M,2004ApJ...612L..25N}, to the Milky Way observed along any line of sight \citep[e.g.,][]{1995A&A...295..655N,2006A&A...451..125V}, to the dwarf galaxies of the Local Group \citep[e.g.,][]{1997RvMA...10...29G,1998ARA&A..36..435M,2009ARA&A..47..371T}, and more distant galaxies as long as their stars can be resolved \citep[e.g.,][]{2016ApJ...823...19C}.
The formalism applies to systems dynamically governed by both collisional and collisionless dynamics.

Finally, a few further notable results of this paper are the following:
\begin{itemize}
	\item We use the mathematical concept of foliation to formalize the multiple-stellar population theory. So far, this is the only known way to reconcile both the time evolution of a distribution function used in classical dynamics theory, and the concept of SSPs typically used in the classical stellar population theory. In this way, we were able to retain the concept of the ''CSP as the sum of SSPs,'' typical for a classical stellar population theory, as well as the time evolution of the distribution functions typically describing classical stellar dynamics. 

	\item We introduced the most general Liouville theorem (see Eq.\eqref{(2.1)bis}) known so far for the conservation of the flux in a given space, and we considered its validity in the space $\mathbb{E}$ defining our concept of CSP. It contains a geometric factor (the metric tensor) that must be accounted for by the general manifold treatment of the existence-space introduced by the composite-stellar population theory. The theory aims to set the basis for a unification of the classical stellar population theory and classical dynamics theory. This equation has no precedent astrophysical use and, therefore, should be of interest to anybody who wants to relate the dynamical ''self-consistency'' with the stellar population ''consistency'' to investigate the distribution function of a time-dependent stellar population.
	
	\item By introducing a ''torii-function'' (see Eq.\eqref{ToryF}) we present the first fully analytical formulation of a polygonal function for an arbitrary number of segments (see Eq.\eqref{(2.32)}). The fact that the IMF requires three sections is just a particular case of Eq. \eqref{(2.32)}, but it provides a general interpolation function between $n$ arbitrary points.

\end{itemize}

\section*{Acknowledgements}
This work was supported by NASA grant NNX14AF84G. SP thanks J. Kollmeier for the fundamental support in the developing process of this project. We thank E. G. Grebel, P. Zeidler, L. Piovan, and R. Tantalo for their contribution to an earlier version of this paper. We thank the anonymous referee for valuable comments.

\bibliographystyle{elsarticle_harv}
\bibliography{Biblio}

\onecolumn

\appendix
\section{Lemma on the solution existence for the MSP-CT}
\textbf{Lemma A1 [Existence of a solution for the MSP-CT]}. Under the same hypothesis of Sec.\ref{Sec2}, the system of Eq.\eqref{(2.16)} has at least one solution.

\textbf{Proof}: We start by making the following assumptions: ${{I}_{\Xi }}\equiv \int_{{\rm M_{l}}}^{{\rm M_{u}}}{dMM\Xi \left( M \right)}$ and ${{I}_{\Psi }}\equiv \int_{{{t}_{1}}}^{{{t}_{2}}}{dt\Psi \left( t \right)}$. Under the same hypothesis of Sec.\ref{Sec2} (where the same notation is exploited for each of the $\text{N}_p$ CSPs)  we obtain (see Eq.\eqref{(2.9)}):
\begin{equation}\label{(A.2)}
{{\rm M}_{c}}={{\xi }_{0}}{{I}_{\Xi }}{{\psi }_{0}}{{I}_{\Psi }}\Rightarrow {{\xi }_{0}}\left( {{\psi }_{0}} \right)=\frac{{{\rm M}_{c}}}{{{I}_{\Xi }}{{\psi }_{0}}{{I}_{\Psi }}}.
\end{equation}
We can determine the star formation constant ${{\xi }_{0}}$ by requiring that the total mass in Eq.\eqref{(2.14)}, \[{{\rm M}_{\text{tot}}}=\sum\limits_{c=1}^{{\text{N}_p}}{\int_{{}}^{{}}{d\bm{x}{{\rho }_{c}}\left( \bm{x} \right)}}\] is achieved after integration. That is: 
\begin{equation}\label{(A.3)}
\sum\limits_{c=1}^{{\text{N}_p}}{\int_{{}}^{{}}{\psi \left( t \right){{\xi }_{c}}\left( {{\rm M}_{c}},{{\xi }_{0}} \right)MdMdt}}={{\rm M}_{\text{tot}}},
\end{equation}
where we explicitly wrote the dependence of the ${{\xi }_{c}}$ on the mass ${{\rm M}_{c}}$ of each CSP and the normalization constant ${{\xi }_{0}}$. Hence, the general setting for the whole of the CSPs reads evidently:
\begin{equation}\label{(A.4)}
\begin{aligned}
& {{\rm M}_{\text{tot}}}=\sum\limits_{c=1}^{{\text{N}_p}}{\int_{{{t}_{1}}}^{{{t}_{2}}}{dt{{\psi }_{c}}\left( t \right)}\int_{{\rm M_{l}}}^{{\rm M_{u}}}{dMM{{\xi }_{c}}\left( M \right)}} \\ 
& ={{\xi }_{0}}\sum\limits_{c=1}^{{\text{N}_p}}{{{\psi }_{0,c}}{{I}_{\Psi ,c}}{{I}_{\Xi ,c}}},  
\end{aligned}
\end{equation}
where for every CSP we assumed that Eqs.\eqref{(2.9)} and \eqref{(2.10)} hold. The previous equation readily yields:
\begin{equation}\label{(A.5)}
{{\xi }_{0}}=\frac{{{\rm M}_{\text{tot}}}}{\sum\limits_{c=1}^{{\text{N}_p}}{{{\psi }_{0,c}}{{I}_{\Psi ,c}}{{I}_{\Xi ,c}}}}.
\end{equation}
Finally, the coefficients ${{\psi }_{0,c}}$ will be obtained from the solution of the system of equations (see Eq.\eqref{(A.2)} and Eq.\eqref{(A.5)}):
\begin{equation}\label{(A.6)}
\begin{matrix}
{{\xi }_{0}}{{\psi }_{0,c}}{{I}_{\Psi ,c}}{{I}_{\Xi ,c}}={{\rm M}_{c}} & \forall c=1,..,\text{N}_p  \\
\end{matrix}.
\end{equation}
It is simple to prove the existence of a solution by observing that the previous system of equations Eq.\eqref{(A.6)} reads:
\begin{equation}\label{(A.7)}\begin{matrix}
\frac{{{\rm M}_{\text{tot}}}}{\sum\limits_{l=1}^{{\text{N}_p}}{{{\psi }_{0,l}}{{I}_{\Psi ,l}}{{I}_{\Xi ,l}}}}{{\psi }_{0,c}}{{I}_{\Psi ,c}}{{I}_{\Xi ,c}}={{\rm M}_{c}} & \forall c=1,..,\text{N}_p,  \\
\end{matrix} 
\end{equation}
that simplifies as
\begin{equation}
\begin{matrix}
\sum\limits_{l=1}^{{\text{N}_p}}{\left( {{\delta }_{cl}}-\frac{{{\rm M}_{c}}}{{{\rm M}_{\text{tot}}}} \right){{I}_{\Psi ,l}}{{I}_{\Xi ,l}}{{\psi }_{0,l}}}=0 & \forall c=1,..,\text{N}_p,  \\
\end{matrix} \\ 
\end{equation}
with $\delta$ the Kronecker's delta; but because
\begin{equation}\label{(A.8)}
\begin{aligned}
& \det \left( \left( {{\delta }_{cl}}-\frac{{{\rm M}_{c}}}{{{\rm M}_{\text{tot}}}} \right){{I}_{\Psi ,l}}{{I}_{\Xi ,l}} \right)=\frac{1}{\text{N}_p!}\sum\limits_{\begin{smallmatrix} 
	{{i}_{1}},{{i}_{2}},...,{{i}_{\text{N}_p}}=1 \\ 
	{{j}_{1}},{{j}_{2}},...,{{j}_{\text{N}_p}}=1 
	\end{smallmatrix}}^{\text{N}_p}{{{\varepsilon }_{{{i}_{1}}...{{i}_{\text{N}_p}}}}{{\varepsilon }_{{{j}_{1}}...{{j}_{\text{N}_p}}}}\left( {{\delta }_{{{i}_{1}}{{j}_{1}}}}-\frac{{{\rm M}_{{{i}_{1}}}}}{{{\rm M}_{\text{tot}}}} \right){{I}_{\Psi ,{{j}_{1}}}}{{I}_{\Xi ,{{j}_{1}}}}\cdot \cdot \cdot \left( {{\delta }_{{{i}_{\text{N}_p}}{{j}_{\text{N}_p}}}}-\frac{{{\rm M}_{{{i}_{\text{N}_p}}}}}{{{\rm M}_{\text{tot}}}} \right){{I}_{\Psi ,{{j}_{\text{N}_p}}}}{{I}_{\Xi ,{{j}_{\text{N}_p}}}}} \\ 
& =\frac{1}{\text{N}_p!}\prod\limits_{i=1}^{\text{N}_p}{\text{N}_p!{{I}_{\Psi ,i}}{{I}_{\Xi ,i}}}\left( {{\rm M}_{\text{tot}}}-\sum\limits_{j}^{{}}{{{\rm M}_{j}}} \right) \\ 
& =0,  
\end{aligned}
\end{equation}
with $\varepsilon $ being the Levi-Civita symbol, infinite solutions exist (in addition to the trivial one) as soon as the constraint ${{\rm M}_{\text{tot}}}-\sum\limits_{j=1}^{\text{N}_p}{{{\rm M}_{j}}}=0$ is satisfied (i.e., always $\forall \text{N}_p$), which concludes the proof.

\end{document}